# MATURE-HEALTH: HEALTH Recommender System for MAndatory FeaTURE choices

Ritu Shandilya, Sugam Sharma, Johnny Wong
Iowa State University, Ames, Iowa
*ritusha@iastate.edu, sugamsha@iastate.edu, wong@iastate.edu*

## Abstract

Balancing electrolytes is utmost important and essential for appropriate functioning of organs in human body as electrolytes imbalance can be an indication of the development of underlying pathophysiology. Efficient monitoring of electrolytes imbalance not only can increase the chances of early detection of disease, but also prevents the further deterioration of the health by strictly following nutrient controlled diet for balancing the electrolytes post disease detection. In this research, a recommender system MATURE-Health is proposed and implemented, which predicts the imbalance of mandatory electrolytes and other substances presented in blood and recommends the food items with the balanced nutrients to avoid occurrence of the electrolytes imbalance. The proposed model takes user's most recent laboratory results and daily food intake into account to predict the electrolytes imbalance. MATURE-Health relies on MATURE-Food algorithm to recommend food items as latter recommends only those food items that satisfy all mandatory nutrient requirements while also considering user's past food preferences. To validate the proposed method, particularly sodium, potassium, and BUN levels have been predicted with prediction algorithm, Random Forest, for dialysis patients using their laboratory reports history and daily food intake. And, the proposed model demonstrates 99.53%, 96.94% and 95.35% accuracy for Sodium, Potassium, and BUN respectively.

MATURE-Health is a novel health recommender system that implements machine learning models to predict the imbalance of mandatory electrolytes and other substances in the blood and recommends the food items which contain the required amount of the nutrients that prevent or at least reduce the risk of the electrolytes imbalance.

## 1. Introduction

The level of electrolytes, metabolic parameters and acid-base are crucial for the human body and an imbalance of these parameters implies the development of an underlying pathology. And an imbalance of electrolytes and acid-base may indicate the underlying pathophysiology such as renal failure, respiratory failure, shock, etc. The laboratory indices are reliable indicators of electrolytes, metabolic parameters and acid-base and depict the cellular and organ function; for example, a significant variation in the laboratory indices such as serum urea nitrogen, creatinine, aspartate aminotransferase (AST), alanine aminotransferase (ALT), and bilirubin are sign of kidney and liver injury or failure of these organs. To better facilitate the patients, these laboratory indices can be used to predict the trend of important electrolytes, metabolic parameters, and acid-base for early detection of patients at risk for deterioration (Gunnerson et al. (2003), Lier et al. (2008)). The symptoms of imbalance electrolytes, metabolic parameters and acid-base and further deterioration of health status are in general imprecise and do not appear until patient's condition worsens and complications occur (Balcı, (2013)). Also, the frequency of laboratory test, even for the patients, suffering prolong chronic illness is low unfortunately as laboratory tests are invasive and expensive; for example, for dialysis patients, insurance companies recommend only one laboratory test per month until there are visible changes in health status. Which makes it difficult to diagnose further deterioration of health status in a timely manner until symptoms become apparently visible due to the worsening of the health status.

Therefore, there is a need of a robust health management system, which monitors and predicts the trend in users' laboratory indices for early detection of imbalanced electrolytes, metabolic parameters and acid-base and helps prevention of further deterioration of health status by addressing the problem in due time (Boroujeni et al. (2019), Redfern et al. (2018), Bauer et al. (2010)). And this research work addresses these two specific problems: 1) predicting the trend in users' laboratory indices for early detection of imbalance of electrolytes, metabolic parameters and acid-base, 2) aiding prevention of further deterioration of health status.

In literature, various machine learning approaches have been proposed to predict the future trend using previous data. These prediction methodologies enable healthcare systems to analyze the future trend of health status of a patient, which allows them to take preventive measures to reduce the risk of further health deterioration. To analyze the trends, the existing algorithms use the data, which is generated by continuous monitoring of health check via performing frequent laboratory tests for regular time intervals, when patients were hospitalized. Given state-of-the-art, there is no existing monitoring device for homes except Continuous Glucose Monitor (CGM), which could do continuous monitoring of electrolytes, metabolic parameters and acid-base at home. However, CGM just monitors blood glucose only and none other than it. Therefore, the needs are imperative to implement methods that predict the future trend of electrolytes, metabolic parameters and acid-base using data from the less frequent non-continuous laboratory test results.

To address this problem, in our research, we propose a machine learning algorithm to predict the imbalance of the electrolytes, metabolic parameters and acid-base. The algorithm is trained on the dataset, which is created by the data from laboratory test performed at irregular time intervals. To test the performance of the proposed algorithm, a dataset generated by the data of the dialysis patient(s) from the University of Iowa Hospital & Clinics (UIHC); the data includes both the inpatient and outpatient laboratory test reports.

Further, after the detection of the possible risk of electrolytes, metabolic parameters and acid-base imbalance, the next step is to take necessary time-bound actions to prevent further deterioration of patients' health. Numerous scientific studies show that diet has been known for many years to play a key role as a risk factor for chronic diseases (Who et al. (2003), Hawkes et al. (2006)) and a monitored nutrient balanced diet may decrease the risk of human-health deterioration. Therefore, various disease specific food recommender systems have been proposed in literature to recommend the diet, which is based on current health status. These food recommender systems are static in nature in the sense that they only consider the current health status of a user and recommend the food according to static food nutritional requirements based on specific disease and do not consider the future trend of individual's health status. Hence, there is a swift need for such a food recommendation mechanism that computes the daily nutritional requirements of a patient dynamically, based on predicted health status after analyzing the future trend of the of electrolytes, metabolic parameters, and acid-base.

Therefore, to address this problem of predicting the trend in users' laboratory indices for early detection of imbalance of electrolytes, metabolic parameters and acid-base and to prevent further deterioration of health status, this research has implemented a health recommender system also, and is dubbed as MATURE-Health and Figure 1 depicts its overview. MATURE-Health ingests - i) patient's laboratory test results, ii) daily nutrient intake record and, iii) electrolytes, metabolic parameters, and acid-base, to be predicted, as input values to the system and builds a feature set, which becomes input to the prediction model. We use Random Forest regression model to predict the future trend of electrolytes, metabolic parameters and acid-base. And after analyzing these future trends, patients' daily nutritional requirements are updated according to the predicted risk of imbalance of metabolic parameters and acid-base. MATURE-Health uses our previously developed MATURE-Food model to make food recommendations using patients' updated daily nutritional requirements and recommends only those food items which fulfill all mandatory nutrient requirements.

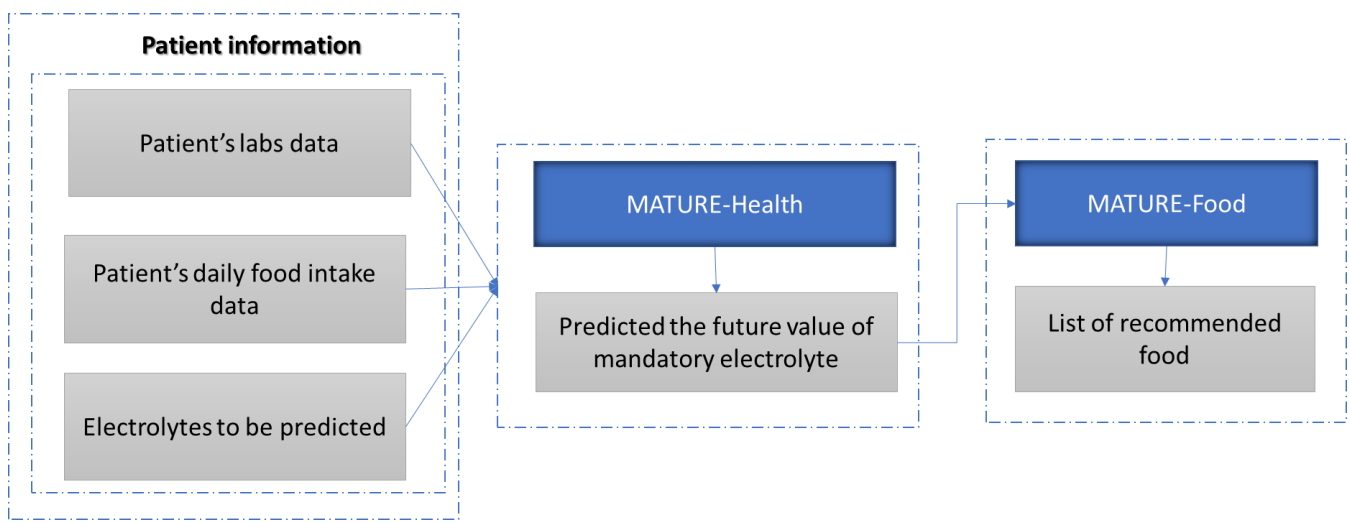

Fig 1: MATURE- Health overview.

As mentioned earlier, in this research the dataset, created by the data, gathered for dialysis patients at UIHC, has been used. Further, three individual machine learning models have been trained to predict the future trend for sodium, potassium and BUN using the history of patient's laboratory test and daily food nutritional intake. And the daily nutritional requirements for sodium, potassium and protein are updated according to the predicted value of sodium, potassium and BUN respectively.

## 2. Related Work

Our MATURE-Health recommender system (also interchangeably called algorithm or model in this paper) is developed and matured to perform two main tasks: i) predicting the health status of the patient in terms of electrolytes based on past labs and daily nutrient intake, ii) applying MATURE-Food model for recommending food items with mandatory nutrients based on predicted health status.

Therefore, this section is focused on the recent research work done in the field of health recommender system (HRS) considering both - predicting the health status, and food recommender system for diet related diseases.

### 2.1 Health recommender system with health status prediction

In the past decade, the exponential advancement in the field of machine learning has given a new perspective to define the problems in different research areas. For the HRS, one of the crucial problems is to predict the health status of an individual based on their past health history and current health status and researchers have made a significant progress to address this problem by using various machine learning algorithms. As this research specifically focuses on diet related diseases, the following section describes the research work addressing the problem of predicting the health status of a user, based on user's lab reports and daily food intake.

According to Houge et al. (2020), diabetes is a leading cause of mortality. The study shows that globally, one of the most significant factors, contributing to the diabetes death, and disability-adjusted life-years is high BMI. Therefore, to reduce the effect caused by diabetes, the researchers are working towards strategies for diabetes prevention and treatment. As for a diabetic patient, analyzing the glucose level trend and predicting the future value of glucose level can prevent from further declining of the health conditions and helps to improve the quality of life; significant efforts have been made to monitoring and analyzing the glucose level trend and predicting the future glucose level. Particularly, CGM technology allows the patient to continuously monitor glucose level in real time and these observations are used to train various type of glucose prediction models to make future glucose level prediction to avoid hyperglycaemia (Mouri et al. (2021)) and hypoglycaemia (Andersen et al. (2020)). Sun et al. (2018) proposed a deep neural network model consisting of one long-short-term memory (LSTM) layer, one bidirectional LSTM layer and several fully connected layers using four different prediction horizons (15, 30, 45, 60) of glucose level. Zhu et al. (2018) proposed a causal dilated convolutional neural network by shifting the outputs by several data points using glucose levels, insulin events, carbohydrate intake and time index. Li et al. (2019) generated an RNN using LSTM, which used glucose readings, insulin bolus, and meal (carbohydrate). They used changes of blood glucose values between the current time, say x(t) and the future, say x(t + 6) as target labels instead of a static blood glucose value. Rabby et al. (2021) proposed a Stacked LSTM based deep recurrent neural network with Kalman smoothing for blood glucose prediction and CGM data, carbohydrates from the meal, bolus insulin, and cumulative step counts in a fixed time interval, are used as input to the model. Woldaregay et al. (2019) published a review of machine learning applications in glucose prediction. Their study shows that the input to train a model mainly consists of blood glucose, change in blood glucose insulin, diet, exercise and stress where CGM technology is used to observe the blood glucose level of user several times a day. But other than CGM, there is still no technology available for continuous monitoring of the electrolytes and other health-related substances like potassium, sodium, phosphorus etc., in blood other than testing the blood sample at laboratory. Monitoring of those electrolytes and other substances is critically important for diseases like chronic kidney disease (CKD) as one of the major impacts of CKD is imbalance of electrolytes and other substances in blood. The following section provides an overview of the recent research in particularly predicting electrolytes, metabolic parameters, and acid-base imbalance. Kwon et al. (2021) developed a deep learning model using electrocardiography (ECG) for detecting the occurrence of electrolyte imbalance, but did not predict the future trend for specific electrolytes, metabolic parameters, or acid-base. Wickramasinghe et al. (2017) and Maurya et al. (2019) predicted the potassium level (Safe, Caution, Danger, noData) using the dataset for CKD patients, which was obtained from the UCI data repository (Asuncion et al. (2007)); there were 400 instances with 25 different attributes related to CKD and the potassium level was predicted using class (ckd, notckd) and 24 attributes, however, they did not predict the future trend for potassium while recommending a diet plan.

## 2.2 Food recommender systems

Ge et al. (2015) showed that the fast and busy lifestyle along with massive availability of food variety had impacted the food choices extensively. And making a healthy food choice has become a critical problem. The studies by Afshin et al. (2019) showed the potential impact of suboptimal diet on non-communicable diseases mortality and morbidity. Therefore, continuous efforts have been made to apply recommender systems in the health/dietary domain to help both end-users/patients and medical professionals. The following section explores some of the important work done in the diet recommender systems to address these issues.

Agapito et al. (2018) and Mata et al. (2018) presented the web-based recommender systems that built user's health profiles based on medical information retrieved from the user/patient, requiring the entry of different laboratory records and vital parameters; and accordingly, provided individualized nutritional recommendations without giving much consideration to the past preferences of the user. Bianchini et al. (2017) proposed PREFer, a food recommender system to provide a personalized and healthy menu, considering both user's preferences and medical prescriptions. However, PREFer did not provide any definitive nutritional information. Rehman et al. (2017) proposed a cloud-based food recommender system, DIET-RIGHT, which was based on user's laboratory reports. This system also highlighted the relevance of selected diets. Tran et al. (2018, 2020), Trang et al. (2018), Trattner et al. (2017), Stark et al. (2019) and Pincay et al. (2019) presented an overview of recommender systems in the healthy food and diet domain and evaluated the existing state-of-the-art diet recommender systems and discussed important contributions, challenges and future research directions related to the development of future diet recommendation technologies; they all inferred that retrieving user information and food nutritional facts was itself challenging; and based on the user information, recommending the correct nutrient value based diet was another level of challenge. From the research work in health recommender systems, following major problems have been identified:

- Continuous monitoring of health status where technology like CGM is not available
- Predicting health status changes based on non-continuous monitoring
- Preventing or reducing the further deterioration of health by taking majors in a timely manner

To address these problems, in this research, we propose a health recommender system (MATURE-Health), which predicts the health status of an individual patient in terms of electrolytes level using their daily food intake and most recent lab reports. And based on these predictions, it subsequently adjusts the nutritional intake and exploits MATURE-Food algorithm to make food recommendations to avoid further health deterioration.

## 3. PROPOSED WORK

This section is composed of two core tasks:
1) Predict the 1-step feature value for mandatory electrolytes, which is further divided into following sub-tasks:

- Compute input feature set and finding electrolytes, which are to be predicted
- Predict 1-step future value
- Update patient's optimized daily nutrient requirements Table

2) Recommend food items based on updated patient's optimized daily nutrient requirements.

Figure 2 is the detailed architecture of our MATURE-Health model, which depicts the above stated each individual task, which are further illustrated here.

## 3.1 Identifying electrolytes and other substances to be predicted

To maintain the health status for an individual, it is necessary to properly keep track of the level of certain electrolytes and other important substances in blood. For example, for a diabetic patient, it is mandatory to keep track of blood glucose level and for a CKD patient the levels of potassium, sodium, BUN etc., need to be monitored. In order to maintain a good health status, this research proposes a preventive approach by predicting the future level of electrolytes and other important substances in a user's blood.

Let $\mathcal{ME}_u^{\mathcal{F}}$ be the set of mandatory electrolytes and other important substance for a user u as represented by Eq (1).

$$\mathcal{ME}_u^{\mathcal{F}} = \{< me_1, min, max >, ... < me_k, min, max >$$

Where, $mn_k =$
$$\begin{cases} 1 & \text{If feature } \mathcal{F}_k \in \mathcal{MN}_u \text{ and} \\ & AI, MI \neq ND \\ 0 & otherwise \end{cases}$$

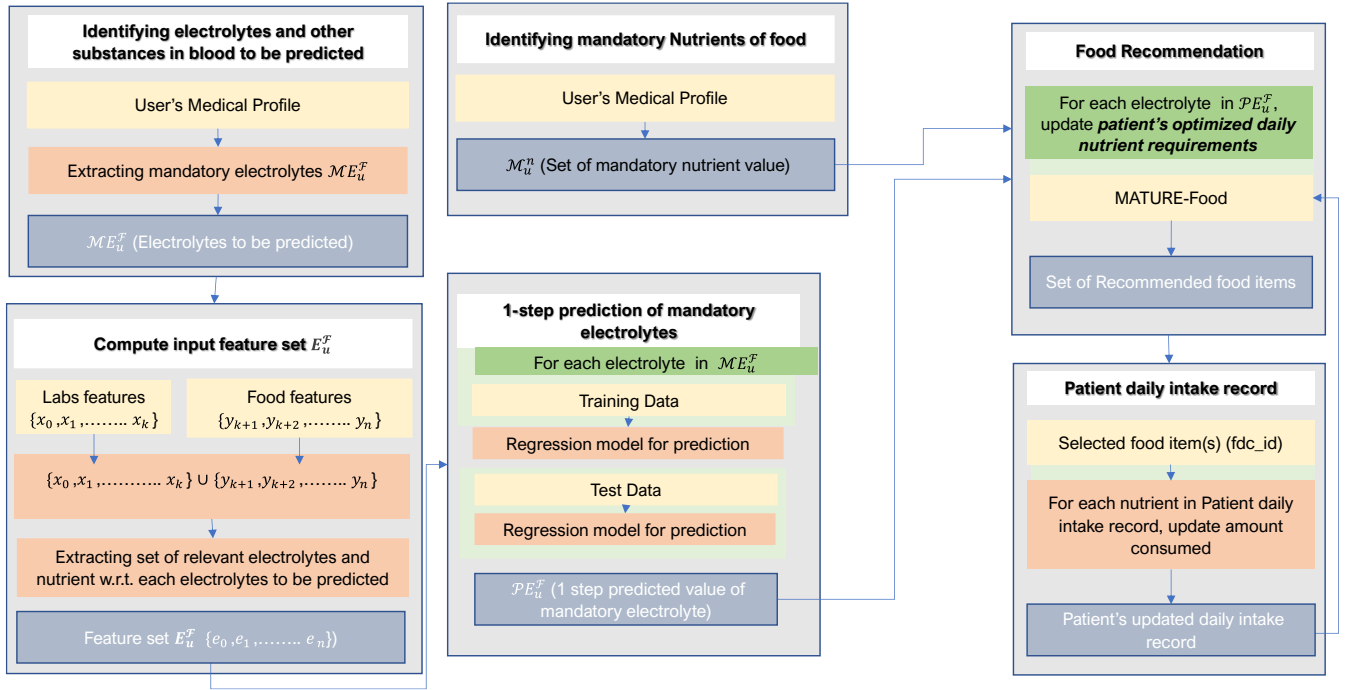

Fig 2: MATURE- Health Architecture. The model is primarily constituted by six modules – Identifying electrolytes and other substances to be predicted, Identifying Mandatory Nutrients, Compute the input set, Predict the 1-step feature value for mandatory electrolytes $\mathcal{PE}_u^{\mathcal{F}}$, Update patient's optimized daily mandatory nutrient requirements, Make Food Recommendations

$, ... < me_p, min, max >\}$            … (1)

Where, $me_k$ is a mandatory electrolytes or other important substances whose future level needs to be predicted to maintain the required health status and p is the total number of those electrolytes or other important substance.

### 3.2 Identifying mandatory nutrients (MNu)

The set of mandatory nutrients for a user u is given by a set $\mathcal{MN}_u$ and $\mathcal{M}_u^n$ contains all the mandatory nutrients and their value range. Each element of $\mathcal{M}_u^n$ is denoted by a tuple $< mn_1, AI, MI >$ where AI is adequate intake, MI is maximum intake for the nutrient $mn_1$. Eq (2) shows vector representation of $\mathcal{M}_u^n$.

$$\mathcal{M}_u^n = \{mn_1 < IA, MI >, ... mn_k < IA, MI >, ... mn_q < IA, MI >\}$$
            … (2)

$\mathcal{MN}_u$ for a user $u$ can be extracted from user's medical profile. For example, for a diabetic patient,

it is mandatory to keep track of blood glucose level, which is highly associated with sugar and carbohydrates in the food intake. For a $u$ with CKD, prescribed daily nutrients intake such as sodium and potassium intake ranges are the source to obtain $\mathcal{MN}_u$. Additionally, the $\mathcal{MN}_u$ can also be derived by ontological mapping based on the other features on the *user's* present profile. For example, for a patient, diabetes on the profile can be mapped to {low sugar, no sugar}.

### 3.3 Compute the input set

To capture the impact of user's daily nutrient intake on user's health, based on user's current health status, MATURE-Health incorporates history of user's laboratory reports and daily food intake and combined these two into an input vector to predict the value of $\mathcal{ME}_u^{\mathcal{F}}$.

#### 3.3.1 Lab Features Extraction

Let's assume that $Lab_u$ is the set of all the labs reports *s.t.* $Lab_u = \{ \mathcal{L}_i \mid i = 1, 2, \ldots, r \}$ where $n = |\mathcal{L}_u|$, represents the total number of past lab reports within a specified timespan $\tau$. Let $\mathcal{F}_{\mathcal{L}}$ be the combined feature set of all the lab results in the set $Lab_u$ s.t. $\mathcal{F}_{\mathcal{L}} = \{ \mathcal{F}_{\ell} \mid j = 1, 2, \ldots, s \}$, the feature set extracted from patent's lab results $m = |\mathcal{F}_{\mathcal{L}}|$ represents the total number of results included in a lab report.

$\forall \mathcal{L}_i \in Lab_u$, a lab report $\mathcal{L}_i$ is represented by a vector in the Eq (3).

$$\mathcal{L}_i = < l_1, l_2, \ldots l_k, \ldots, l_s > \qquad \ldots (3)$$

Where, $l_k$ = result value of $\mathcal{F}_k$ , i = 1 to r and k =1 to s.

### 3.3.2 Food Item Features Extraction

The proposed work utilizes the food item's nutritional information, which constitutes the formation of the item's feature set. Let's assume that $\mathcal{J}$ is a set of all the food items s.t. $\mathcal{J} = \{ \mathcal{J}_i \mid i = 1, 2, \ldots, n \}$ where $n = |\mathcal{J}|$, the cardinality of the $\mathcal{J}$ represents the total number of food items in set $\mathcal{J}$. Let $\mathcal{F}_{\mathcal{J}}$ be the combined feature set of all the items in the set $\mathcal{J}$ s.t. $\mathcal{F}_{\mathcal{J}} = \{ \mathcal{F}_j \mid j = 1, 2, \ldots, m \}$, the food item's feature set extracted from nutritional information; $m = |\mathcal{F}|$.

$\forall \mathcal{J}_i \in \mathcal{J}$, any item $\mathcal{J}_i$ is represented by given as a vector in Eq (4).

$$\mathcal{J}_i = < f_1, f_2, \ldots f_k, \ldots, f_m > \qquad \ldots (4)$$

Where, $f_k$ = {value for feature $\mathcal{F}_k$ in $\mathcal{J}_i$ and {i = 1 to n and k =1 to m}.

### 3.3.3 Compute Input Feature Vector $\mathbb{I}_u$

To predict the future value of electrolyte or other substance $me_k$ in the blood, it is crucial to find all the electrolytes or substances in blood, which influence the level of $me_k$. For each $me_k$ in $\mathcal{ME}_u^{\mathcal{F}}$, let $fl_u(me_k)$ be the set of all the electrolyte or substance which are associated with $me_k$ and $fi_u(me_k)$ is the set of all the food nutrients, which influence the level of $me_k$ in the blood. Let $E_u^{\mathcal{F}}$ be the set of features that jointly represents the daily food intake and most recent lab results of a user u s.t.

$$E_u^{\mathcal{F}}(me_k) = \{ fi_u(me_k) \cup fl_u(me_k)\} \qquad \ldots (4)$$

where $fi_u(me_k) \subseteq \mathcal{F}_{\mathcal{J}}$ and $fl_u \subseteq \mathcal{F}_{\mathcal{L}}$ }

Let $\mathbb{I}_u$ be the input set for a user u where $\forall \mathbb{I}_u(i) \in \mathbb{I}_u$, $\mathbb{I}_u(i)$ is represented by the feature vector with same feature set $E_u^{\mathcal{F}}$.

### 3.4 Predict the 1-step feature value for mandatory electrolytes $\mathcal{PE}_u^{\mathcal{F}}$

After identifying electrolytes and other substances, to be predicted, as set $\mathcal{ME}_u^{\mathcal{F}}$ and computing input set $\mathbb{I}_u$, next step is to predict the next future value of all the $me_k$ in $\mathcal{ME}_u^{\mathcal{F}}$ using prediction algorithm described in experimentation section. For each $me_k$ in $\mathcal{ME}_u^{\mathcal{F}}$, let $\mathcal{PE}_u^{\mathcal{F}}$ be the set of predicted future value

$$\mathcal{PE}_u^{\mathcal{F}} = \{ pe_1, pe_2, \ldots pe_k, \ldots pe_p \} \qquad \ldots (5)$$

where $pe_k$ is the next predicted value of $me_k$.
Following is the self-descriptive pseudocode for proposed Prediction Algorithm.

**Input:** Patient's daily nutrient intake set $fid_u$ , patient's lab history $Lab_u$ mandatory nutrient value set $\mathcal{M}_u^n$ , mandatory electrolytes value set $\mathcal{ME}_u$ , input feature set $E_u$

**Prediction Algorithm**

```
1   Initialization:  𝕀u ← ∅, Eu𝓕 ← ∅, ℳEu𝓕 ← ∅, 𝒫Eu𝓕 ← ∅
2   Compute input feature set Eu𝓕 (fidu, Labu) :
3        𝕀u ← fidu  ∪ Labu
4        Eu𝓕 ← extract_feature_set(𝕀u)
5   End
6   Return (Eu𝓕)

7   Finding electrolytes to be predicted (ℳu𝑛):
8        For each mn k ∈ ℳun
9            For each ek ∈ Eu𝓕
10               If mn k is relevant ek:
11                   add ek and corresponding value range of mnk to set ℳEu𝓕
12               End
13           End
14       End
15   Return (ℳEu𝓕) // electrolytes to be predicted

16  Predict 1-step future value (ℳEu𝓕):
17       For each mek ∈ ℳEu𝓕
18           Prediction algorithm (mek, 𝕀u)
19           Add (1-step future value for mek to 𝒫Eu𝓕 )
20       End
21   Return (𝒫Eu𝓕)

22  Update patient's optimized daily nutrient requirements Table :
23       For each pek ∈ 𝒫Eu𝓕 :
24           If (pek > max_mek  or  < min_mek):
25               Update mn k range accordingly in patient's optimized
26               daily nutrient  requirements Table..
27           End
28       End
29   Return(optimized daily nutrient requirements Table)
```

### 3.5 Update patient's optimized daily mandatory nutrient requirements

If $pe_k$, the next predicted value of $me_k$ does not lie within the specified range, the intake value of each mandatory nutrient $mn_k$ belongs $fi_u(me_k)$ should be updated as $fi_u(me_k)$, the set of all the food nutrients, which influence the level of $me_k$. Let the optimized daily mandatory nutrient requirements of a user is represented by a vector $\mathcal{OM}_u^n$ as shown in Eq (6)

$\mathcal{OM}_u^n = \{on_1 < IA, MI >, \dots on_k < IA, MI >$
$, \dots on_q < IA, MI >\}$ ...... (6)
Where, $\forall\ on_k \in \mathcal{OM}_u^n\ \ on_k = mn_k$

Update $on_k$ range accordingly in patient's optimized daily nutrient requirements vector as shown in Eq (7)

$\forall\ pe_k \in \mathcal{PE}_u^{\mathcal{F}}$,

$on_k < IA, MI > =$
$\begin{cases} on_1 < MI > - on_1 < MI > \% \rho\ \ If\ (pe_k > \max\_me_k) \\ on_1 < IA > + on_1 < IA > \% \rho\ \ If\ (pe_k < \min\_me_k) \\ on_k < IA, MI > = on_k < IA, MI > \quad \quad otherwise \end{cases}$
$\dots (7)$

### 3.6 Make food recommendations

After computing the optimized daily food intake for a user, based on predicted values of mandatory electrolytes and other important substance in $\mathcal{ME}_u^{\mathcal{F}}$, we used MATURE-Food (Shandilya et al. (2022)) algorithm for constituting the food recommendations as MATURE-Food ensures to fulfill all mandatory nutrient requirements along with respecting the user's past preferences, while recommending the food items. The following section gives an overview of the MATURE-Food algorithm.

#### 3.6.1 MATURE-Food for MATURE-Health

The optimized daily mandatory nutrient requirements of a user, represented by a vector $\mathcal{OM}_u^n$ are new mandatory nutrient requirements for user u. Following are the simple steps for proposed Recommendation Algorithm. Additionally, some relevant content of MATURE-Food algorithm is also mentioned here again to help readership with easy, better and wholistic understanding of MATURE-Health.

***Recommendation Algorithm***

| | |
|---|---|
| **Input:** | Seed training set T, classification strategy Sc, food item set $\mathcal{I}$, optimized mandatory nutrient set $\mathcal{M}_u^n$, set of food items liked by user u $\mathcal{IU}_u$, meal # |

**Food Recommendation Algorithm (**Seed training set T, classification strategy Sc, food item set $\mathcal{I}$, optimized mandatory nutrient set $\mathcal{M}_u^n$, set of food items liked by user u $\mathcal{IU}_u$, meal **#)**

1   Mature-food (T, Sc, $\mathcal{I}$,$M_u^n$, $\mathcal{IU}$ )
2   Return ($\mathcal{R}_u$)

#### 3.6.2 MATURE-Food algorithm

First, MATURE-Food classifies the food items $\in \mathcal{I}$ into $\mathcal{C}$ classes based on the nutritional similarity and for each class $\mathcal{C}_x$, a class weight vector $\overrightarrow{\mathbb{CWV}_{\mathcal{C}_x}}$ which represents the nutritional value range of the food items belonging to class $\mathcal{C}_x$.

For each user u, vector $\overrightarrow{\mathbb{PWV}_u}$ represents the historical preferences and a vector $\mathcal{M}_u^{\mathcal{F}}$ denotes the mandatory requirements of the user. To capture the user's past preferences and current mandatory requirements together, the combined feature vector, $\overrightarrow{\mathbb{UWV}_u}$ is computed.

Algorithm further finds the set $\mathcal{C}_{\mathcal{U}}^{\mathcal{T}}$, the set of top k classes containing food items most similar to user's past preferences along with satisfying all mandatory requirements using Eq (8) for user $u$.

$\mathcal{S}_{\mathcal{W}}(\overrightarrow{\mathbb{CWV}_{\mathcal{C}_x}}, \overrightarrow{\mathbb{UWV}_u}\ )$
$= \dfrac{(\overrightarrow{\mathbb{CWV}_{\mathcal{C}_x}} \cdot \overrightarrow{\mathbb{UWV}_u})}{(||\overrightarrow{\mathbb{CWV}_{\mathcal{C}_x}}||_2\ ||\overrightarrow{\mathbb{UWV}_u}||_2)}$
$\dots (8)$

For all the food items belongs to any of the classes in the set $\mathcal{C}_{\mathcal{U}}^{\mathcal{T}}$, algorithm computes similarity between food item $\overrightarrow{\mathcal{I}_{\mathcal{S}}}$ and $\overrightarrow{\mathbb{PWV}_u}$ as in Eq (9) and the requirement satisfaction score, $\mathcal{RqS}_{\mathcal{W}}$ as Eq (10).

$\mathcal{S}_{\mathcal{W}}(\overrightarrow{\mathcal{I}_{\mathcal{S}}}, \overrightarrow{\mathbb{PWV}_u})\ =\ \dfrac{(\overrightarrow{\mathcal{I}_{\mathcal{S}}} \cdot \overrightarrow{\mathbb{PWV}_u})}{(||\overrightarrow{\mathcal{I}_{\mathcal{S}}}||_2\ ||\overrightarrow{\mathbb{PWV}_u}||_2)} \qquad \dots (9)$

$$\mathcal{R}q\mathcal{S}_W(\overrightarrow{\mathcal{I}_S}, \overrightarrow{\mathrm{UWV}_u})$$

$$= \begin{cases} \dfrac{(\overrightarrow{\mathcal{I}_S} \cdot \overrightarrow{\mathrm{UWV}_u})}{\| \mathcal{P}_u^{\mathcal{F}} \| + \epsilon} + 1 & if \ m\mathfrak{f}_k = 1 \ for \ \mathcal{M}_u^{\mathcal{F}} \ and \ \mathfrak{f}_k = 1 \ for \ \mathcal{I}_S \\[12pt] \dfrac{(\overrightarrow{\mathcal{I}_S} \cdot \overrightarrow{\mathrm{UWV}_u})}{\| \mathcal{P}_u^{\mathcal{F}} \| + \epsilon} & Otherwise \end{cases}$$

$$\dots \ (10)$$

Algorithm recommends top k most similar food items based on similarity score $\mathcal{S}_W(\overrightarrow{\mathcal{I}_S}, \overrightarrow{\mathbb{PWV}_u})$ if $\mathcal{R}q\mathcal{S}_W(\overrightarrow{\mathcal{I}_S}, \overrightarrow{\mathrm{UWV}_u}) \geq 1$.

## 4. Experimental Procedure and Results

### 4.1 Dataset

The dialysis unit of University of Iowa Stead Family Children's Hospital at University of Iowa Hospitals & Clinics (UIHC) has medical data of CKD patients and for this research, we obtained the data from there. The dataset has lab records of End-Stage Renal Disease (ESRD) patients including CBC W Diff, Basic Metabolic Panel, Phosphorus, Magnesium, Albumin, Alkaline Phosphatase, Vitamin D 25 Hydroxy, Uric Acid. The time span of the data collected ranges from the beginning of dialysis till the time when a kidney is transplanted. We along with patient's lab records, also used the records of patient's daily food intake including water intake, as balancing fluid intake is crucial for CKD patients. The daily food intake is also used to determine the preferences of food items for a specified time interval. Any limitation imposed on daily nutrient intake is also retrieved from the patient's medical records as mandatory nutrient requirements.

In this research, for food recommendation, a publicly available dataset at USDA website: U.S. DEPARTMENT OF AGRICULTURE "the Food and Nutrient Database for Dietary Studies" - (https://fdc.nal.usda.gov/download-datasets.html) has been used.

### 4.2 Methodology

#### 4.2.1 Identifying electrolytes and other substances to be predicted

As the dataset used in this study has records of (ESRD) patients at dialysis unit, the experimentation for this research focuses on End-Stage Renal Disease (ESRD). The major role of kidney is to maintain acid-base and electrolyte homeostasis in human body which has high impact on various metabolic processes and organ functions in the human body but decline in kidney function leads to acid-base and electrolyte imbalance in human body (Van et al. (2020)). Therefore, nephrologists rely on pathological labs to monitor the acid-base and electrolyte homeostasis in the human body and control the food nutrient intake to achieve the best possible acid-base and electrolyte homeostasis.

This research proposes to predict the next day value of the most important electrolytes and other substances for CKD patients. Table 1 shows the standard value range of most important electrolytes and other substances for CKD patients (Dhondup et.al. (2017), Commonwealth Nephrology Associates). For this experimentation, sodium, potassium and BUN have been considered as mandatory electrolytes and other substances, which are to be predicted. Table 1 shows all the mandatory electrolytes and other substances belongs to $\mathcal{ME}_u^{\mathcal{F}}$.

**Table 1**

Standard Range electrolytes and other substances

| Component | Standard Range |
|-----------|----------------|
| Sodium | 135 - 145 mEq/L |
| Potassium | 3.5 - 5.0 mEq/L |
| Chloride | 95 - 107 mEq/L |
| Phosphate | 2.5 - 4.5 mg/dL |
| Magnesium | 1.5 - 2.9 mg/dL |
| Calcium | 8.5 - 10.5 mg/dL |
| CO2 | 22 - 29 mEq/L |
| BUN | 10 - 20 mg/dL* |
| Creatinine | 0.5 - 1.0 mg/dL |

#### 4.2.2 Identifying mandatory nutrients (MNu)

A set of high impact nutrients for CKD patients is selected (Please refer, Nutrition for Children with Chronic Kidney Disease[1]) and based on daily nutrient requirements, for each nutrient, a default value is set (Please refer, Dietary Guidelines for Americans; Dietary Guidelines for Americans [2] ; Nutrition Education Resources & Materials [3] ; The National Center for Biotechnology Information[4]). Table 2 shows the default daily nutrient requirements for healthy children. Patient's medical record is used to identify patient's daily mandatory dietary requirements and the values of nutrients intake is updated based on those mandatory requirements. Vector $\mathcal{M}_u^{\mathcal{F}}$ is updated as shown in Eq (3). Table 3 shows an example of nutrient values consumed daily user $u1$.

---

[1] https://www.niddk.nih.gov/health-information/kidney-disease/children/caring-child-kidney-disease/nutrition-chronic-kidney-disease

[2] https://www.dietaryguidelines.gov/

[3] https://www.fda.gov/food/food-labeling-nutrition/nutrition-education-resources-materials

[4] https://www.ncbi.nlm.nih.gov/

### 4.2.3 Compute the input set

In this research, $me_k$ s belongs to $ME_u^\mathcal{F}$ are sodium, potassium and BUN and, the future value of each of these items is predicted based on the most recent lab results and user's daily food intake record.

#### 4.2.3.1 Lab features extraction

In general, for CKD patients, the level of sodium, potassium and BUN is monitored by blood tests, which are performed monthly or as needed based on patient's health status. Practically, these tests are not as frequent and regular as CGM until patient is hospitalized. Therefore, for predicting the $pe_k$ corresponding to each of sodium, potassium and BUN, most recent blood test results are used. $\mathcal{F}_\mathcal{L}$ shows the set of electrolytes and other substances considered as lab features to build input feature set.

$\mathcal{F}_\mathcal{L}$ = {A_Gap, Calcium, Chloride, CO2, Creatinine, Potassium, Sodium, Phosphorus, Bun}

#### 4.2.3.2 Food item's nutrient extraction

The abnormality of electrolytes and other substances in blood is the result of CKD and to maintain the balance of those electrolytes and other substances in blood, it is necessary to observe and control the intake of nutrients which influence the level of respective electrolytes and other substances. For example, BUN is highly influenced by the intake of protein and can be controlled by increasing or decreasing the amount of protein consumed daily. The set of nutrients, which influence the value of Sodium, Potassium and BUN is represented by $\mathcal{F}_j$.

$\mathcal{F}_j$ = {Beneprotein, Renvela, sodium polystyrene sulfonate powder, potassium chloride, Food-Potassium, Food-Protein, Food-Sodium, Food-Phosphorus, water}

Table 2
Dietary Intakes Reference Values for nutrient

| Age | Chloride (mg/d) AI | MI | Iron (mg/d) AI | MI | Phosphorus (mg/d) AI | MI | Potassium (mg/d) AI | MI | Sodium (mg/d) AI | MI | Total Protein g/kg/day AI | MI | Total Water Liter/day AI | MI |
|---|---|---|---|---|---|---|---|---|---|---|---|---|---|---|
| **Infants** | | | | | | | | | | | | | | |
| 0-6 mo | 180 | ND | 0.27 | 40 | 100 | ND | 400 | ND | 120 | ND | 1.52 | ND | 0.7 | ND |
| 7-12 mo | 570 | ND | **11** | 40 | 275 | ND | 700 | ND | 370 | ND | 1.2 | ND | 0.8 | ND |
| **Children** | | | | | | | | | | | | | | |
| 1-3 y | 1500 | 2300 | **7** | 40 | 460 | 3000 | 3000 | ND | 1000 | 1500 | 1.05 | ND | 1.3 | ND |
| 4-8 y | 1900 | 2900 | **10** | 40 | 500 | 3000 | 3800 | ND | 1200 | 1900 | 0.95 | ND | 1.7 | ND |
| **Males** | | | | | | | | | | | | | | |
| 9-13 y | 2300 | 3400 | **8** | 40 | 1250 | 4000 | 4500 | ND | 1500 | 2200 | 0.95 | ND | 2.4 | ND |
| 14-18 y | 2300 | 3600 | **11** | 45 | 1250 | 4000 | 4700 | ND | 1500 | 2300 | 0.85 | ND | 3.3 | ND |
| **Females** | | | | | | | | | | | | | | |
| 9-13 y | 2300 | 3400 | **8** | 40 | 1250 | 4000 | 4500 | ND | 1500 | 2200 | 0.95 | ND | 2.1 | ND |
| 14-18 y | 2300 | 3600 | **15 e** | 45 | 1250 | 4000 | 4700 | ND | 1500 | 2300 | 0.85 | ND | 2.3 | ND |

Source: Health Canada: http://www.hc-sc.gc.ca/fn-an/alt_formats/hpfb-dgpsa/pdf/nutrition/dri_tables-eng.pdf
Abbreviation: AI: Adequate Intake, MI: maximum intake, ND: not determined
For any nutrient in MI, ND indicates that non-availability of data for this study and does not infer that there is no upper limit for reference intake.

Table 3
Patient daily nutrient requirements

| Age | Chloride (mg/d) AI | MI | Iron (mg/d) AI | MI | Phosphorus (mg/d) AI | MI | Potassium (mg/d) AI | MI | Sodium (mg/d) AI | MI | Total Protein g/day AI | MI | Total Water Liter/day AI | MI |
|---|---|---|---|---|---|---|---|---|---|---|---|---|---|---|
| 5 y | NM | NM | 8 | NM | NM | 368 | NM | 700 | NM | NM | 38 | NM | 1.3 | 1.5 |

Source: Patient's health record
Abbreviation: AI: Adequate Intake, MI: maximum intake, ND: not determined, NM: Not mandatory implies no specific value for the user
Intake values are based on formula intake though formula was treated to control Phosphorus and Potassium level. Sodium polystyrene sulfonate powder is used to control potassium. Renvela (sevelamer carbonate) is used to control phosphorus levels.

*4.2.3.3 Compute Input Feature Vector $\mathbb{I}_u$*

After extracting the lab features $\mathcal{F}_L$ and food nutrients as food features $\mathcal{F}_j$ for sodium, potassium and BUN, the combined input feature sets $E_u^{\mathcal{F}}(me_k)$ are extracted as shown in Eq (4). From Eq (11) to Eq (19), we show $fi_u(me_k)$, $fl_u(me_k)$ and $E_u^{\mathcal{F}}(me_k)$ for sodium, potassium, and BUN respectively.

$fi_u$(Sodium) = {sodium polystyrene sulfonate powder, potassium chloride, Food-Sodium, water} … (11)

$fl_u$(Sodium) = {A_Gap, Calcium, Chloride, CO2, Creatinine, Potassium, Sodium, Phosphorus, Bun} … (12)

$E_u^{\mathcal{F}}$(Sodium) = {sodium polystyrene sulfonate powder, potassium chloride, Food-Sodium, water, Sodium, A_Gap, Calcium, Chloride, CO2, Creatinine, Potassium, Sodium, Phosphorus, Bun} … (13)

$fi_u$(Potassium ) = {sodium polystyrene sulfonate powder, potassium chloride, Food-Potassium, water} … (14)

$fl_u$(Potassium )= {A_Gap, Calcium, Chloride, CO2, Creatinine, Potassium, Sodium, Phosphorus, Bun} … (15)

$E_u^{\mathcal{F}}$(Potassium) = {sodium polystyrene sulfonate powder, potassium chloride, Food-Potassium, water, A_Gap, Calcium, Chloride, CO2, Creatinine, Potassium, Sodium, Phosphorus, Bun} … (16)

$fi_u$(BUN ) = {Beneprotein, Food-Protein, water} … (17)

$fl_u$(BUN ) = {A_Gap, Calcium, Chloride, CO2, Creatinine, Potassium, Sodium, Phosphorus, Bun} … (18)

$E_u^{\mathcal{F}}$(BUN ) = {Beneprotein, Food-Protein, water, A_Gap, Calcium, Chloride, CO2, Creatinine, Potassium, Sodium, Phosphorus, Bun} … (19)

### 4.2.4 Predict the 1-step feature value for mandatory electrolytes $\mathcal{P}E_u^{\mathcal{F}}$

To predict the 1-step feature value for each mandatory electrolyte $me_k$ belongs to $\mathcal{M}E_u^{\mathcal{F}}$, three independent Random Forest models have been trained for each $me_k$, sodium, potassium, and BUN. To find the best prediction models, three different prediction models, Random Forest, LSTM (Long and Short-Term Memory Network), and XGBoost have been evaluated for each of sodium, potassium, and BUN and compared based on the evaluation criteria explained in section 4.4. The models trained with Random Forest outperform the other models. In this research, the next day's levels of sodium, potassium, and BUN are predicted based on the last 3 days intake of nutrients and laboratories results. To make those predictions, Random Forests have been implemented as a time series forecasting method. And in order to do so, first, the time series dataset has been transformed into supervised learning problem dataset by restructuring it. Let dataset for each $me_k$ is represented as D = {($x_i, y_i$ ), i=1,…,n} where $x_i = (x_{i1,……,} x_{ir})$ and r = $|E_u^{\mathcal{F}}(me_k)|$. To restructure the dataset for random forest, 3 consecutive rows of D are combined by applying sliding window method to create each row of new dataset which is represented by

$D_{rf}$ = {( ($x_{i+0}, x_{i+1}, x_{i+2}$), $y_{i+2}$ ), i=1,…,n}. The grid search approach has been used to train a model for each of sodium, potassium, and BUN using 60% test-train split and 5-fold cross validation. Trained model for sodium, potassium, and BUN used to predict the next day's level of sodium, potassium, and BUN respectively.

### 4.3 Update patient's optimized daily mandatory nutrient requirements

The prediction of the next day's level of sodium, potassium, and BUN directs the optimization of the patient's daily nutrient requirements. For predicted blood level of sodium, the intake amount of sodium, for potassium, the intake amount of potassium, and for BUN, the intake amount of protein is required to be updated. Table 1 is used as a reference for different electrolytes and other substances blood level range. The patient's daily nutrient requirements are optimized according to Eq (7) as according to Dr. Lyndsy Harshman at University of Iowa, starting approach should be to change the intake value by 10%.

### 4.4 Making food recommendations

Once patient's daily mandatory nutrient requirements are optimized, food recommendations are made using MATURE-Food algorithm that ensures to recommend the only food items, which have nutrient level within the specified range. In order to maintain the history of patient's daily intake and to keep the track of nutrients consumption, patient's daily nutrient consumption record is used. Table 4 depicts an instance of a patient's daily nutrient consumption record; this record reveals the amount of each nutrient consumed up to a specific time of the day. This information is utilized to make food recommendations as described in section 3.6.2.

**Table 4**
Patient daily nutrient consumption record

| Meal | Iron (mg/d) | Phosphorus (mg/d) | Potassium (mg/d) | Sodium (mg/d) | Total Protein g/day | Total Water Liter/day |
|------|-------------|-------------------|------------------|---------------|---------------------|-----------------------|
| 1 | 2.31 | 187 | 188 | 522 | 10.81 | 0 |
| 2 | 0 | 101 | 150 | 38 | 3.28 | 88 |
| 3 | 2.66 | 100 | 215 | 398 | 9.51 | 0 |
| 4 | 0 | 0 | 0 | 0 | 0 | 0.5 |

### 4.5 *Results*

This section demonstrates the evaluation criteria used to evaluate the performance of the implemented models and the comparison of results. As this research predicts the 1-step future value of sodium, potassium and BUN, Table 5 shows the value range for sodium, potassium and BUN for the dataset.

**Table 5**
Value range for electrolytes to be predicted

| | Sodium | Potassium | BUN |
|--|--------|-----------|-----|
| **Hypo-Value** | 127 | 2.3 | 28 |
| **Mean-Value** | 136.12 | 3.86 | 79.89 |
| **Hyper-Value** | 145 | 6.2 | 110 |

### 4.4.1 Evaluation criteria

The results of research done by Chicco et al. demonstrate that the coefficient of determination $R^2$ is more informative and truthful than symmetric mean absolute percentage error (SMAPE), MAE, MAPE, MSE and RMSE in regression analysis evaluation. Therefore, in this research, to evaluate the performance of models, Accuracy, mean absolute percentage error (MAPE), mean absolute error (MAE), mean square error (MSE), root mean square error (RMSE), and coefficient of determination or R-squared ($R^2$) are used as evaluation metrics (Botchkarev et al.).

The formula for calculating the MAE is as follows:

$$MAE = \frac{1}{n} \sum_{i=1}^{N} |\hat{y}_i - y_i|$$

The formula to calculate MAPE is as follows:

$$MAPE = \frac{1}{n} \sum_{i=1}^{N} \frac{|\hat{y}_i - y_i|}{y_i} \times 100\%$$

The formula for calculating the MSE is as follows:

$$MSE = \frac{1}{n} \sum_{i=1}^{N} (y_i - \hat{y}_i)^2$$

The formula to calculate RMSE is as follows:

$$RMSE = \sqrt{\frac{1}{n} \sum_{i=1}^{N} (y_i - \hat{y}_i)^2}$$

There are several variants, $R^2$ can be presented. The most widely used form to compute the value of $R^2$ is represented by the following formula:

$$R^2 = 1 - \frac{\text{Sum squared regression (SSR)}}{\text{Total sum of squares (SST)}}$$

$$\text{Or} \qquad R^2 = 1 - \frac{\sum(y_i - \hat{y}_i)^2}{\sum(y_i - \overline{y})^2}$$

Where, N is the total number of data points, $y_i$ is actual value, $\hat{y}_i$ is predicted value and $\overline{y} = \frac{1}{n} \sum_{i=1}^{N} y_i$.

### 4.4.2 Results for sodium

Table 6 shows the prediction results for sodium using Random Forest and LSTM. The value range for potassium is between 127 and 145 with an average value of 136.12. MSE for sodium is 39.731 which is a significantly low *w.r.t* value range of sodium.

**Table 6**
**Sodium**

| | Random Forest | LSTM |
|--|---------------|------|
| Accuracy | 99.53% | 98.64% |
| MAPE | 0.004 | 0.013 |
| MAE | 0.643 | 1.84 |
| MSE | 2.324 | 4.84 |
| RMSE | 1.290 | 2.200 |
| R-squared | 0.849 | 0.370 |

Figure 3 visualizes the comparison of actual sodium level to predicted sodium level for Random Forest.

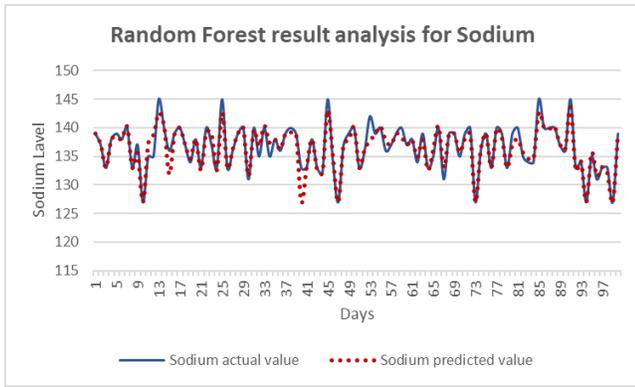

Fig 3: Plot for sodium using Random Forest

Figure 4 depicts the comparison of actual sodium level to predicted sodium level for LSTM.

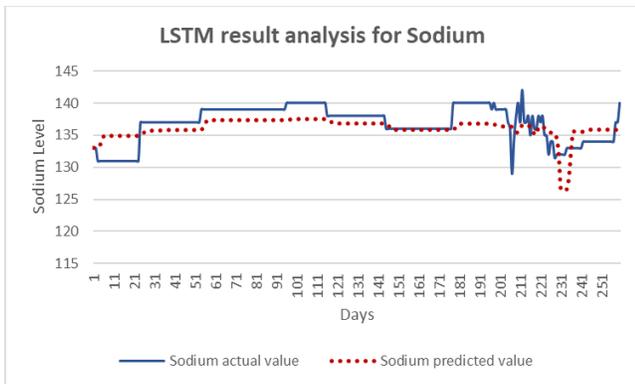

Fig 4: Plot for sodium using LSTM

### 4.4.3 Results for potassium

Table 7 manifests the prediction results for the potassium using Random Forest and LSTM. The value range for potassium is between 2.3 and 6.2 with an average value of 3.86. MSE for potassium is 0.080, which is a low *w.r.t* value range of potassium and accuracy for Random Forest is 96.94%.

**Table 7**
**Potassium**

|  | Random Forest | LSTM |
| --- | --- | --- |
| Accuracy | 96.94% | 92.25% |
| MAPE | 0.030 | 0.077 |
| MAE | 0.110 | 0.362 |
| MSE | 0.080 | 0.601 |
| RMSE | 0.283 | 0.775 |
| R-squared | 0.782 | 0.246 |

Figure 5 shows the comparison of actual potassium level to predicted potassium level for Random Forest.

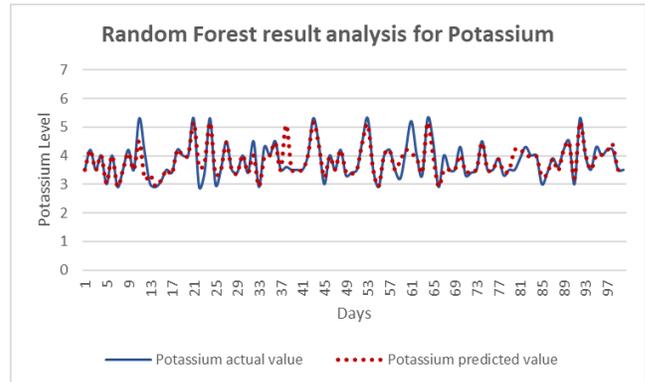

Fig 5: Plot for potassium using Random Forest

Figure 6 exhibits the comparison of actual potassium level to predicted potassium level for LSTM.

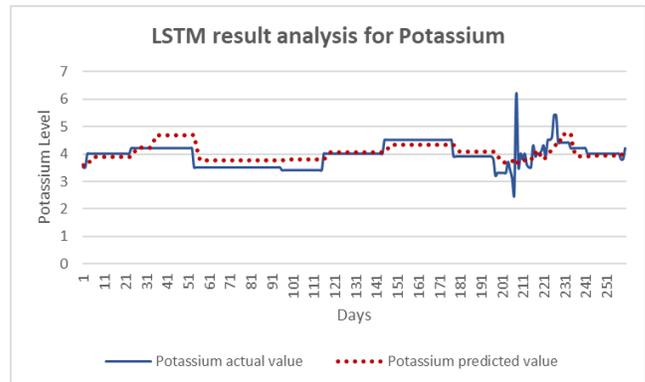

Fig 6: Plot for potassium using LSTM

### 4.4.4 Results for BUN

Table 8 shows the prediction results for the BUN using Random Forest and LSTM. The value range for BUN is between 28 and 110 with an average value of 79.89. MSE for BUN is 39.731 and accuracy for Random Forest is 95.35%.

RF Model parameters:

**Table 8**
**Bun**

|  | Random Forest | LSTM |
| --- | --- | --- |
| Accuracy | 95.35%. | 82.14%. |
| MAPE | 0.046 | 0.178 |
| MAE | 2.859 | 13.631 |
| MSE | 39.731 | 244.942 |
| RMSE | 6.303 | 15.650 |
| R-squared | 0.901 | 0.119 |

Figure 7 visualizes the comparison of actual BUN level to predicted BUN level for Random Forest.

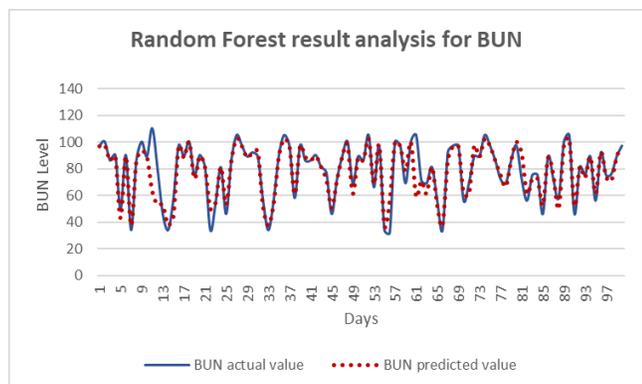

Fig 7: Plot for BUN using Random Forest

Figure 8 demonstrates the comparison of actual BUN level to predicted BUN level for LSTM.

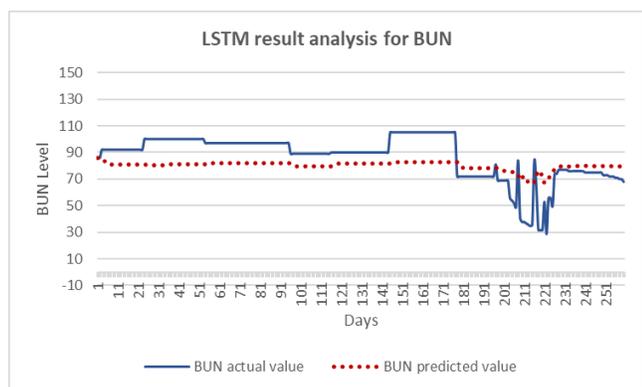

Fig 8: Plot for BUN using LSTM

The results, shown in Table 6, 7 & 8 validate that Random Forest perform better than LAST for our dataset. In case of sodium, the MAE and MSE for Random Forest is marginally lower than the LSTM but $R^2$ value for Random Forest (0.849) is significantly higher in comparison to LSTM (0.37). As for a regression model, $R^2$ demonstrates the integrity of fit of a model. It is a statistical evaluation measure that explains how perfectly the regression line estimates the actual data, the performance of Random Forest is substantially better than LSTM. For values range of potassium, the difference between Randon Forest and LSTM is considerably high for MAE, MSE and RMSE respectively. But $R^2$ for LSTM is drastically lower than Random Forest. In comparison of the other two models (sodium and potassium), the performance of LSTM for BUN is worst. Not only MAE and MSE are severely high, but the value of $R^2$ is radically low (.011) in comparison to Random Forest (0.901). Therefore, for predicting the 1-step future value of all three of sodium, potassium and BUM, Random Forest outperforms LSTM.

## 5. CONCLUSION AND FUTURE WORK

In this research, we proposed and developed a health recommender system, called MATURE-Health, in which we investigated numerous machine learning algorithms to predict the abnormalities in the level of electrolytes and other substances present in blood. And based on the predicted level of electrolytes and other substances, MATURE-Health provided the individualized nutritional recommendation, which determined daily food requirements for a user. The food items are recommended according to the nutrient controlled daily food requirements to prevent the occurrence of electrolytes imbalance and to reduce the risk of further deterioration of health. This research utilized user's medical record to identify the electrolytes and other substances that needed to be predicted. User's recent lab history and daily food intake data were used to train the models to predict electrolytes imbalance. Our MATURE-Health used the Random Forest model to predict the level of sodium, potassium and BUN present in blood and provides reliable predictions. Our previously developed MATURE-Food algorithm was also used to make food recommendations using daily food requirements, which were determined by the predicted level of sodium, potassium and BUN as mandatory nutrient requirements. Our proposed methodologies could be employed to get better insights into CKD patient's future electrolyte level that can lead to a best health management system as electrolytes imbalance can be prevented by providing a controlled nutrients intake.

As future work, MATURE-Health is intended to be extended further to provide other more critical disease specific models to predict the electrolytes imbalance.

### ACKNOWLEDGEMENTS

Authors would like to acknowledge Dr. Lyndsy Harshman, Associate Professor of Pediatrics-Nephrology, Dialysis and Transplantation, and Rachel Beiler, Dietitian at UIHC for their continuous guidance and support about understanding the problem, accessing data and other related discussion for several years.